\documentstyle[12pt]{article}
\begin{document}
\setlength{\baselineskip}{3.0ex}

\title{{\it B} Decays - Measurements and Predictions
\footnote{A talk given at 'B  Physics at Hadron Colliders', Snowmass 1993.}}
\author{Hitoshi Yamamoto \\
        {\it Harvard University} \\
        {\it 42 Oxford St., Cambridge, MA 02138, USA} }
\date{  }
\maketitle

\begin{abstract}
Hadronic decays of $B$ mesons are reviewed. First, masses of $B$ mesons
and observed patterns together with physics behind them are
discussed. Then the effective Hamiltonian responsible for major decays is
presented and its practical applications are discussed in the context of
factorization. Various tests of factorization are then studied. For rare
decays, the focus is placed on $K\pi$, $\pi\pi$ final state and the
penguin-mediated $X_s\gamma$. In general, the measurements are in excellent
agreement with predictions of the standard model.
\end{abstract}

\section{Basic Methods on Upsilon-4S Resonance}

Most of the data presented in the following are collected on the
upsilon-4S resonance, and some basic experimental techniques are briefly
described below.

The B meson pair production cross section on the upsilon-4S resonance is
roughly 1 nb; namely, an integrated luminosity of 1 fb$^{-1}$ would
generate 1 million B meson pairs. The CLEO-II detector has logged about
1.2 fb$^{-1}$ of data thus generating 1.2 million B meson pairs.

On the upsilon-4S resonance, light quark pairs ($u \overline u$, $d
\overline d$, $s \overline s$, and $c \overline c$ - often referred to as
the `continuum') are also generated in
addition to the B meson pairs. The cross section ratio of B meson pair
to the continuum is roughly 1 to 2.5. The continuum is often a major
background and in order to understand this component, data are taken right
below the resonance (32 MeV below the peak) corresponding to about one
half of the integrated luminosity taken on the resonance. When we want to
plot a distribution of certain parameter for B meson pairs, we can subtract
the distribution for the data taken off-resonance from that taken
on-resonance (with a proper normalization). The distribution is then said
to be `continuum subtracted'.

At the upsilon-4S resonance, the B mesons are generated with definite
energy and momentum given by
\begin{equation}
   E_B = E_{\rm beam} = 5.289 {\rm GeV}, \qquad P_B = 0.325 {\rm GeV/c}
\end{equation}
When reconstructing a decay $B \to f_1 + f_2 + \cdots f_n$, natural
parameters to look at are thus the total energy and momentum of the decay
products $f_i$ $(i=1,..,n)$:
\begin{equation}
   E_{\rm tot} = \sum E_i, \qquad P_{\rm tot} = \sum \vec P_i
\end{equation}
which should peak at $E_{\rm beam}$ and $P_B$ respectively,
where $E_i$ and $P_i$ are the energy and momentum of the i-th decay
product. In practice, often used parameters are the `energy difference'
$\Delta E$ and the `beam-constrained mass' $M_B$ defined by
\begin{equation}
  \Delta E = E_{\rm tot} - E_{\rm beam}, \qquad
   M_B = \sqrt{E_{\rm beam}^2 - P_{\rm tot}^2}.
  \label{eq:MBdE}
\end{equation}
Since $E_{\rm beam}$ is a constant, measuring $\Delta E$ and $M_B$ is
equivalent to measuring $E_{\rm tot}$ and $P_{\rm tot}$. The mass
reconstructed this way has a good resolution which varies from 2.5
to 3.3 MeV depending on decay mode and usually dominated by the spread of
beam energy. The essence of this method in background rejection,
however, is
simply the conservation of energy and absolute momentum in a B meson decay.
We will often be referring to $M_B$ and $\Delta E$ in the rest of this
article; the definitions are as defined above.

\section{Masses}

\subsection{${\it B}^-$ {\it and} $\overline {\it B}^0$}

The masses of neutral and charged B mesons can be measured by fully
reconstructing the major decay modes. Figure~\ref{fg:Bmasses} shows the
distribution of the beam-constrained mass $M_B$ for $B^-$ and $\overline
B^0$ mesons after requiring that the energy difference $\Delta E$ is within
$2.5\sigma$ of zero. The decay modes used are $B^-\to D^{*0}\pi^-,
D^{*0}\rho-,D^0\pi^-,D^0\rho^-,\psi K^-$ for the charged $B$ meson and
$\overline B^0\to D^{*+}\pi^-,D^{*+}\rho^-,D^+\pi^-, D+\rho^-,\psi K^{*0}$
for the neutral $B$ meson. The $D^*$ mesons are detected by the decays
$D^{*0}\to D^0\pi^0, D^{*+}\to D^0\pi^+$ and $D$ mesons are detected by
$D^0\to K^-\pi^+, D^+\to K^-\pi^+\pi^+$. These modes are chosen since they
are particularly clean. There are 362 signal events for $B^-$ and 340
signal events for $\overline B^0$. With a correction due to initial state
radiation of $-1.1\pm0.5$ MeV, we obtain $M_{B^0} = 5280.3\pm0.2\pm2.0$ MeV
and $M_{B^-} = 5279.9\pm0.2\pm2.0$ MeV . The first error is statistical and
the second systematic. The systematic error is dominated by the uncertainty
in the energy scale of the storage ring which cancels when we take the mass
difference: $M_{B^0} - M_{B^-} = 0.44\pm0.25\pm0.19$ MeV; namely, the
masses of $B^-$ and $\overline B^0$ are consistent with being identical
within several tenth of MeV. The results are summarized in
Table~\ref{tb:masses} together with previous measurements.

\begin{table}
\begin{center}
\begin{tabular}{|c|c|c|c|}
\hline
\hline
    (MeV) & CLEO I.5 \cite{CLEOmasses}
          & ARGUS \cite{ARGUSmasses}
          & CLEO II \cite{CLEOIImasses} \\
\hline
  $M_{B^0}$ & $5278.0\pm0.4\pm2.0$ & $5279.6\pm0.7\pm2.0$
                           & $5280.3\pm0.2\pm2.0$ \\
  $M_{B^-}$ & $5278.3\pm0.4\pm2.0$ & $5280.5\pm1.0\pm2.0$
                           & $5279.9\pm0.2\pm2.0$ \\
  $M_{B^0}-M_{B^-}$ & $-0.4\pm0.6\pm0.5$ & $-0.9\pm1.2\pm0.5$
                           & $0.44\pm0.25\pm0.19$ \\
\hline
\end{tabular}
\end{center}
\caption{Masses of neutral and charged B mesons.}
\label{tb:masses}
\end{table}

It is interesting to compare this result with that for strange and charm
mesons. There we have $M_{K^0}-M_{K^-} = 4,024\pm0.032$ MeV and
$M_{D^+}-M_{D^0} = 4.77\pm0.27$ MeV \cite{PDG} which seem to indicate that
the meson mass is heavier when a heavy quark is combined with a $d$ quark
than with a $u$ quark. The pattern, however, clearly
does not repeat for $B$
mesons. The current understanding for the isospin mass splitting is that
there are effects due to the $u-d$ mass difference as well as QED effects
\cite{Mesonmasses}
(i.e. due to the electric charge difference between $u$ and $d$ quarks).
Both are of order a few MeV,  and the two kinds of effects happen to cancel
for the $B$ meson case \cite{Bsplit}. There seems to be no simple and
intrinsic reason to give $M_B^0 = M_B^+$.

\begin{figure}
  \vspace{3.7in}
  \caption{The beam-constrained mass for charged (a) and neutral (b) B
     mesons after $\Delta E$ is required to be consistent with zero.
     Particularly clean modes are selected and summed.}
  \label{fg:Bmasses}
\end{figure}

\subsection{{\it Other Bottom Mesons}}

Bottom hadrons heavier than $B^-$ and $\overline B^0$ are not produced on
upsilon-4S resonance, and the results so far come from accelerators that
operate at higher energies.

Figure~\ref{fg:CDFBs}
\begin{figure}
  \vspace{3.5in}
  \caption{$B_S\to\psi\phi$ decay observed by the CDF collaboration. (a)
   Invariant mass of $\psi K^+K^-$ when the $K^+K^-$ pair forms the $\phi$
   mass (within $\pm10$ MeV). The dots are for a $\phi$ mass side band. (b)
   Invariant mass of $K^+K^-$ when the $\phi K^+K^-$ mass is in the $B_S$
   peak (within$\pm20$ MeV).}
  \label{fg:CDFBs}
\end{figure}
shows the decay $B_S\to
\psi\phi,\phi\to K^+K^-$ observed by the CDF collaboration \cite{CDFBs} in
$p\overline p$ collisions at 1.8 TeV c.m. energy. There are $14\pm4.7$
events observed and fitting a gaussian to the peak, the $B_S$ mass is
determined to be $5383.3\pm4.5\pm5.0$ MeV. The ALEPH collaboration has
also reported a result on $B_S$ mass from two events $B_S\to \psi'\phi$ and
$D_S^+\pi^-$. The mass measurement is dominated by the $\phi'\phi$ event and
gives $5369\pm5.6\pm1.5$ MeV. These results are summarized in
Table~\ref{tb:Bsmass} together with a possible candidate event reported
earlier by the OPAL collaboration and a recently reported result from
DELPHI.
\begin{table}
\begin{center}
\begin{tabular}{|c|c|c|c|}
\hline
\hline
          & Modes & Number of events
          & $M_{B_S}$ (MeV) \\
\hline
  ALEPH \cite{ALEPHBs} & $\psi'\phi,D_S^+\pi^-$
                       & $2$
                       & $5368.6\pm5.6\pm1.5$ \\
  CDF \cite{CDFBs} & $\psi\phi$
                   & $14\pm4.7$
                   & $5383\pm4.5\pm5.0$ \\
  OPAL\cite{OPALBs} & $\psi\phi$
                   & (1 candidate)
                   & $5360\pm70$ \\
  DELPHI\cite{DELPHIBs} & $D_S^+(\pi^- {\rm or} a_1^-), \psi\phi$
                   & 4
                   & $5357\pm12\pm6$ \\
  \hline
  \end{tabular}
  \end{center}
  \caption{Measurements of $B_S$ meson mass. The $\psi^{(')}$ and
   $\phi$ mesons are detected by $\psi^{(')} \to l^+l^-$ and $\phi\to
   K^+K^-$ respectively, and $D_s^+$ mesons are detected in the modes
   $D_S^+\to \phi\pi^+,K^*K$. }  \label{tb:Bsmass}
 \end{table}
The measurements by CDF and ALEPH are marginally consistent (2-sigma
difference statistically); taking the weighted average, the mass difference
between $B_S$ and $B^{0}$ is 97 MeV. The value is strikingly similar to
the charm case $M_{D_S^+}-M_{D^+} = 99.5\pm0.6$ MeV \cite{PDG}, and also
consistent with predictions of non-relativistic models: $M_{B_S}=
5345-5388$ MeV \cite{Bmassth}.

The mass of $B^*(J^P = 1^-)$ has been measured by CUSP \cite{CUSPB*} and
CLEO \cite{CLEOB*} by detecting the monochromatic photon in the transition
$B^*\to B\gamma$. The numbers are \cite{B*corr}
\begin{eqnarray}
     M_{B^*} - M_B =& 46.4\pm0.3\pm0.8 \quad &\hbox{MeV (CLEO)} \\
                   & 45.6\pm1.0             &\hbox{MeV (CUSP)}.
\end{eqnarray}
These measurements are in accordance with an intriguing observation on
the hyperfine splitting
\begin{equation}
    \Delta M\equiv M^2(1^-) - M^2(0^-) = \hbox{const}\approx 0.5
    \hbox{GeV}^2.   \label{eq:Msqdif}
\end{equation}
This holds well for $(\pi,\rho)$, $K$, $D$, $D_S$ and now for $B$. In
non-relativistic models, such relation is realized when the potential
between the constituent quarks is linearly increasing as a function of the
distance between the quarks \cite{Bmassth,DeltaM2}. It is consistent with
a naive picture that the two constituent quarks are connected by a flux
tube with a constant tension. At short distance, the potential is expected
to be Coulomb-like; this portion of the potential, however, is not
expected to play a significant role \cite{MassesTh}. Also, there is an
electromagnetic hyperfine splitting which violates the relation
\ref{eq:Msqdif}, but its effect is also much smaller than the hyperfine
splitting due to strong interaction \cite{Amundson}.

Apart from the theoretical importance, the above mass difference indicates
that $B^*$ cannot decay to $B\pi$. It has a practical implication that one
cannot tag the sign of the bottom flavor by the decays such as $B^{*+}\to
B^0\pi^+$ where the charge sign of the pion tells us if the neutral $B$
meson is bottom or anti-bottom. Such flavor tagging would have made it easy
to study the CP violating decay asymmetry in $B^0$ or $\overline B^0\to
\psi K_S, \pi^+\pi^-$ etc. particularly in hadron colliders. Now we have to
hope that there may be a higher resonance that decays to $B\pi$ which is
narrow and produced copiously \cite{B**tag}.

\section{Non-Suppressed Decays}

\subsection{{\it Effective Hamiltonian}}

The interaction of interest for $B$ meson decays comes from the charged
current part of the Standard Model Lagrangian \cite{STDMtexts}:
\begin{equation}
    {\cal L}_{CC} = {g\over\sqrt2} (u,c,t)_L\gamma_\mu V
         \left( \begin{array}{c}
                 d \\ s \\ b
                \end{array}
         \right)_L W^\mu.
\end{equation}
where  $g$ is the weak coupling constant, the subscript $L$ for the quark
field indicates left-handed component (e.g. $u_L={1\over2} (1-\gamma_5)u$
etc.), and the matrix $V$ is the Cabbibo-Kobayashi-Masukawa (CKM) matrix:
\begin{equation}
   V \equiv
         \left(
            \begin{array}{ccc}
             V_{ud} & V_{us} & V_{ub} \\
             V_{cd} & V_{cs} & V_{cb} \\
             V_{td} & V_{ts} & V_{tb}
            \end{array}
         \right)   .
\end{equation}
The experimental value of the CKM matrix $V$ is well represented by
\cite{CKMvalues,Wolfen} (assuming unitarity of $V$)
\begin{equation}
   V \sim
         \left(
            \begin{array}{ccc}
             1     & \lambda & |V_{ub}|{\rm e}^{i\alpha} \\
             -\lambda    & 1       & \lambda^2           \\
             |V_{td}|{\rm e}^{i\beta}   & -\lambda^2 & 1
            \end{array}
         \right)
    \qquad {\rm where}\quad
      \left\{
         \begin{array}{l}
             \lambda \sim 0.22 \\
             \alpha = \arg(V_{ub}) \\
             \beta  = \arg(V_{td})
         \end{array}
      \right.
\end{equation}
and the magnitude of $V_{ub}, V_{td}$ is of order $\lambda^3$. Taking the
first and third columns, the unitarity condition
\begin{equation}
V_{ud}^*V_{ub} + V_{cd}^*V_{cb} + V_{td}^*V_{tb} = 0
\end{equation}
becomes a triangle as below (called the unitary triangle).
\begin{equation}
  \vspace{2.5in}
\end{equation}

At energy scales well below the $W$ mass, the propagation of $W$ can be
`integrated out' and we obtain 4-fermion effective Hamiltonian \cite{HeffB}
relevant to $B$ decays given by
\begin{equation}
   {\cal H}_{\rm eff} = {G_F \over \sqrt2} V^*_{ud}V_{cb}
   (C_1(\mu){\bf O}_1 + C_2(\mu){\bf O}_2) + \cdot\cdot\cdot
   \label{eq:Heff}
\end{equation}
\begin{equation}
    {\bf O}_1 = (\overline d  u) (\overline c b), \qquad
    {\bf O}_2 = (\overline c  u) (\overline d b)
\end{equation}
where
$G_F=g^2/(4\sqrt2 M_W^2)$ is the Fermi coupling constant and the quark
current $(\overline q' q)$ is a short hand for $\overline
q'_\alpha\gamma_\mu(1-\gamma_5)q_\alpha$ which is a color-singlet $V-A$
current ($\alpha$: color index).  Any combination of replacements $c\to
u, u\to c$ and $d\to s$ can be made to obtain other possible interactions as
long as the replacements are consistently made including the indexes of the
CKM matrix elements.  The terms shown in (\ref{eq:Heff})
are part of an expansion of the effective hamiltonian (the operator product
expansion \cite{OPE}). It has an advantage that the calculable
short-distance effects are separated into the coefficients of the operators
(Wilson coefficients) while the long distance effects such as
the state of valence quarks in mesons are absorbed into matrix elements of
the operators.

Without QCD correction, we only have the first operator ${\bf O}_1$ which is
shown diagrammatically in Figure~\ref{fg:Heff}(a). With QCD correction,
gluons flying between the quark lines can shuffle the color flows and
generate an effective neutral current operator ${\bf O}_2$ shown in
Figure~\ref{fg:Heff}(b). The Wilson coefficients $C_{1,2}$ can be
calculated using the leading-logarithm approximation (LLA) \cite{LLA}
\begin{equation}
   C_1 = {1\over2}(C_+ + C_-) \qquad
   C_2 = {1\over2}(C_+ - C_-)
\end{equation}
with
\begin{equation}
   C_\pm = \left[
          {\alpha_S (\mu^2) \over \alpha_S(M_W^2)}
           \right]^{d_\pm \over 2b}
\end{equation}
where $d_- = -2d_+ = 8$, and $\alpha_S$ is the running coupling
constant of strong interaction given by
\begin{equation}
    \alpha_S(\mu^2) = {4\pi \over b \log(\mu^2/\Lambda^2_{\rm QCD})}
     \qquad {\rm with} \quad b = 11 - {2\over3}n_f .
\end{equation}
with $n_f$ being the number of relevant flavors, and $\mu$
the typical mass scale of problems in question. Note that $C_+$ and $C_-$
are related by  $C_+^2 C_- = 1$. With $\mu=m_b=5$ GeV, $n_f=4$,  and
$\Lambda_{\rm QCD} = 0.25$ GeV we have
\begin{equation}
     C_1(m_b) = 1.11 \qquad C_2(m_b) = -0.26 . \label{eq:C12val}
\end{equation}
The next-to-leading logarithm approximation (NLLA) has
been computed \cite{NLLA}; the result does not differ drastically from the
LLA result quoted above.
For the transition $b\to cs\overline c$, however, the momentum transfer
associated with the light quarks are much smaller than the bottom mass
scale and as a result the corresponding coefficients could be significantly
different from (\ref{eq:C12val}).  In fact, in one estimation using heavy
quark effective theory (HQET) \cite{HQET}, the coefficients are about 30\%\
larger for $C_1$ and almost twice as large for $C_2$ \cite{Coefbcsc}:
\begin{equation}
     C_1\sim 1.45 \qquad C_2\sim -0.45
      \qquad \hbox{(for $b\to cs\overline c$)}     \label{eq:C12bcsc}
\end{equation}

There are also 4-fermion operators of the type shown in
Figure~\ref{fg:Heff}(c) called Penguin operators \cite{Penguin}. The
corresponding coefficients, however, are small and the Penguin
operators are relevant only for highly suppressed decays such as $B\to
K^*\gamma$ and $K\pi$, to which we will come back later.

\begin{figure}
  \vspace{3.5in}
  \caption{Four fermion operators of the effective Hamiltonian responsible
   for $B$ meson decays.}
  \label{fg:Heff}
\end{figure}

\subsection{{\it Two-body Decays and Factorization}}

Compared to semileptonic decays, hadronic decays are
harder to understand due to variety of short and long-distance strong
interactions among the quarks involved. Two-body hadronic decays, however,
are the simplest kind, and some framework of understanding - factorization
- exists \cite{fact-rev}. Also, it should be noted that two-body decays
account for a substantial fraction of total hadronic decays of heavy mesons
($\sim$ 15\%\ for bottom mesons and $\sim$ 75\%\ for charm mesons when
resonances are included \cite{BSW}).

The idea of factorization for hadronic weak decay dates back at least to
the early 60's when Schwinger showed that the $\Delta I = 3/2$ transition
of $K\to\pi\pi$ can be estimated from the corresponding semileptonic rate
\cite{Schwinger}. The procedure, however, was not considered to be
accurate; in fact, when Feynman reported calculations of $\Lambda\to
p\pi$ and $K^+\to\pi^+\pi^0$ using the idea of factorization
\cite{Feynman}, he preceded the discussion by the following disclaimer:
`You may not wish to consider this line of flimsy reasoning; we are
becoming very uncertain about this matter, nevertheless I shall present
it.' There is, however, a good reason to believe that the factorization
works well for certain $B$ decays.

We take $\overline B^0\to D^+\pi^-$ as an example. This can occur by
the operator ${\bf O}_1$ as shown in Figure~\ref{fg:BDpi} (a),
\begin{figure}
  \vspace{3.5in}
  \caption{Decay $\overline B^0\to D^+\pi^-$ by the operator ${\bf O}_1$ (a)
   and ${\bf O}_2$ (b). The latter is suppressed by a factor $\xi$.}
  \label{fg:BDpi}
\end{figure}
where it is
assumed that the $B\to D$ transition is caused by the current
operator $(\overline c b)$ and that $\pi^-$ is created by the
current operator $(\overline d u)$. Assuming that the $B\to D$ transition
and the $\pi^-$ creation are independent, the amplitude can be written as
\begin{equation}
    \langle D^+\pi^- | (\overline d  u) (\overline c b) |
         \overline B^0 \rangle
     = \langle \pi^- | (\overline d  u) | 0 \rangle
       \langle   D^+ | (\overline c  b) | \overline B^0 \rangle
\end{equation}
which constitutes the essence of the factorization assumption.

It is
instructive to visualize the situation intuitively. A $B$ meson may be
viewed as an analog of a hydrogen atom where the heavy bottom quark is
at the center surrounded by a cloud made of light quark and gluon
[Figure~\ref{fg:BDpifact}(a)].
\begin{figure}
  \vspace{3.5in}
  \caption{An intuitive picture of the decay $\overline B^0\to D^+\pi^-$.
   Before the decay (a), immediately after the $b$ quark decay (b), and
   right after the formation of final state mesons (c).}
  \label{fg:BDpifact}
\end{figure}
Upon the decay of the $b$ quark, the $b$ quark disappears and $c$,
$\overline u$, and $d$ quarks appear. The $c$ quark will combine with the
original cloud that was around the $b$ quark to form a $D$ meson, and the
$\overline u d$ pair will eventually turn into a pion. Here one can cast
doubts on the factorization assumption on two points:
\begin{enumerate}
  \item When the $\overline u d$ pair passes through the cloud, it may
   strongly interact with the cloud, in which case the formation of the
   $D$ meson and the creation of the pion cannot be independent.
  \item After the $D$ meson and the pion are formed, they may re-scatter
   through final-state interaction (FSI); e.g. $D^+ + \pi^- \to D^0 +
   \pi^0$ etc.
\end{enumerate}
For each of the above, Bjorken has argued that it does not pose serious
problem for the factorization assumption \cite{BjorkenFact}. First, the
invariant mass of the $\overline u d$ pair is of order pion mass; thus,
they are highly collinear and close together. Since the total color of the
pair is zero, they form a small color dipole and the cloud cannot see them
from some distance away. The pair is thus expected to pass through the cloud
without much interaction. Second, the formation time of the pion in its own
rest frame is of order $0.3$ fm/c which is the time for light to propagate
from the center of the pion to the edge. Since the pion is highly energetic
($\sim 2.5$ GeV), by the time it is formed the distance between the $D$
meson and the pion is already several fermis; thus, they cannot interact
through FSI. A similar argument of `color transparency' was also used for
production of $\rho$ and $\psi$ in high energy scatterings \cite{Brodsky}.

This line of argument has been put forward by Dugan and Grinstein in the
framework of QCD and the heavy quark effective theory, and it has been
shown that factorization holds in the limit of $M_{B,D}\to\infty$ while
$M_B/M_D$ is kept constant \cite{Dugan-Grinstein}. For decays which involve
two charmed mesons such as $\overline B^0\to D^-_S D^+$, the two mesons in
the final state are partially overlapped at the formation time, and thus the
factorization may not work well for these decays. Factorization is known to
hold also for the large $N_C$ limit where $N_C$ is the number of colors
\cite{Buras1/N}. Even though the correction to the limit is of order
1/3 which is quite large, the applicability of the $1/N_C$ argument is not
restricted to the large velocity limit \cite{Reader-Isgur}, and thus
complementary to the `color transparency' argument.

The decay $\overline B^0\to D^+\pi^-$ can also proceed by the operator
${\bf O}_2$ as shown in Figure~\ref{fg:BDpi}(b). In this case, naively only
the color singlet component of the $\overline u$ and $d$ legs is expected to
contribute. Applying Fierz transformations to color indexes as well as to
gamma matrices \cite{Okun}, ${\bf O}_2$ can be written as
\begin{equation}
    {\bf O}_2 = {1\over3}{\bf O}_1 +
      {1\over2}(\overline d \lambda^i u) (\overline c \lambda_i b)
    \label{eq:O2color}
\end{equation}
where the second term is a color singlet operator formed by two
color-octet currents with $\lambda^i$ being the $SU(3)$ Gell-Mann
matrices. Thus, ${\bf O}_2$ contains ${\bf O}_1$ within itself, and
consequently ${\bf O}_1$ and ${\bf O}_2$ are {\it not} orthogonal
\cite{Reader-Isgur}. The overall coefficient of ${\bf O}_1$ is then
$C_1+C_2/3$. For the decay $\overline B^0\to D^0\pi^0$, the relevant
operator is ${\bf O}_2$. There, the role of ${\bf O}_1$ and ${\bf
O}_2$ are inverted with the overall coefficient of ${\bf O}_2$ being
$C_2+C_1/3$. In fact, we can write (\ref{eq:Heff}) in two ways
\begin{equation}
  \begin{array}{ll}
    C_1{\bf O}_1 + C_2{\bf O}_2
     &= (C_1 + {C_2\over3}) {\bf O}_1 +
        {1\over2}(\overline d \lambda^i u) (\overline c \lambda_i b) \\
     &= (C_2 + {C_1\over3}) {\bf O}_2 +
        {1\over2}(\overline d \lambda^i b) (\overline c \lambda_i u)
   \end{array}  \label{eq:O12Fierz}
\end{equation}
Assuming factorization, the effective Hamiltonian may then be written in
terms of `factorized hadron operators' \cite{fact_oper} as
\begin{equation}
   {\cal H}_{\rm had} = {G_F \over \sqrt2} V^*_{ud}V_{cb}
   [ a_1(\overline d u)_{\rm had} (\overline c b)_{\rm had} +
     a_2(\overline d b)_{\rm had} (\overline c u)_{\rm had} ]
   \label{eq:Hhad}
\end{equation}
where the above arguments suggest
\begin{equation}
    \begin{array}{ll}
      a_1 = & C_1 + \xi C_2 \\
      a_2 = & C_2 + \xi C_1
    \end{array}
    \qquad {\rm with} \quad \xi = {1\over3},   \label{eq:a12}
\end{equation}
where the effect of ${\bf O}_2$ to the first term and that of ${\bf O}_1$
to the second term is parametrized by $\xi$ (sometimes called 'color
suppression factor'). The contribution of the octet current term in
(\ref{eq:O2color}), however, may have a significant effect; in fact, an
estimation using QCD sum rule indicates that its contribution may in effect
lead to $\xi\sim 0$ \cite{Blok-Shifman}. Also, an analysis of charm
decays suggests $\xi$ near zero \cite{BSW}. It has thus been suggested that
$a_1,a_2$ be taken as free parameters \cite{BSW}.

Given the factorized Hamiltonian (\ref{eq:Hhad}), one can then write down
the amplitude for a decay. For example, if $X^-$ is a meson made of
valence quarks $d$ and $\overline u$,
\begin{equation}
   Amp(\overline B^0 \to D^+X^-) =
     {G_F \over \sqrt2} V^*_{ud}V_{cb} a_1
        \langle X^- | (\overline u d)_{\rm had}^\mu | 0 \rangle
     \langle D^+ | (\overline c b)_{{\rm had}\mu} | \overline B^0 \rangle
\end{equation}
where we have from Lorentz invariance
\begin{eqnarray}
\langle X^- | (\overline u d)_{\rm had}^\mu | 0 \rangle
      & = -if_X q^\mu \quad
      & \hbox{(for $X$: pseudo scalar)}\\
\langle X^- | (\overline u d)_{\rm had}^\mu | 0 \rangle
      & = f_X m_X\epsilon^\mu \quad
      & \hbox{(for $X$: vector or axial vector)} \label{eq:dconvec}
\end{eqnarray}
with $f_X$ being a parameter of energy dimension (called the decay
constant). The current matrix element is the same as that appears in the
corresponding semileptonic decay \cite{WSBsemilep} evaluated at
$q^2=m_\pi^2$:
\begin{equation}
     \langle D^+ | (\overline c b)_{{\rm had}\mu} | \overline B^0 \rangle
     = \left( P_B  + P_D - {m_B^2 - m_D^2 \over q^2} q \right)_\mu
         F_1(q^2)
        + {m_B^2 - m_D^2 \over q^2} q_\mu F_0(q^2)
\end{equation}
where $F_0$ and $F_1$ are longitudinal and transverse form factors
respectively [one can easily verify that the coefficient of $F_1$
satisfies $(...)_\mu q^\mu = 0$]. For the case of pion emission, the
transverse component exactly vanishes (by definition) and we have
\begin{equation}
   Amp(\overline B^0 \to D^+\pi^-) =
      -i{G_F \over \sqrt2} V^*_{ud}V_{cb} a_1 f_\pi
     (m_B^2 - m_D^2) F_0(m_\pi^2). \label{eq:BDpiamp}
\end{equation}
The form factors $F_{0,1}$ may be either obtained from semileptonic decays
or calculated by models such as the relativistic harmonic oscillator model
together with the pole dominance \cite{WSBsemilep}. They are relatively
slowly varying functions of order 1. In addition, the heavy quark
effective theory allows us to relate all form factors for transitions
between heavy mesons to a universal form factor \cite{HQETform}. Similar
procedures are applied to other decay modes.

In general, we may distinguish three classes of decays when we consider
two-body decays of heavy mesons mediated by operators of the types ${\bf
O}_{1,2}$ in spectator mode (i.e. the light quark in the parent meson does
not participate in the weak decay) \cite{3classes}:
\begin{description}
   \item[Class 1] Only the first term in (\ref{eq:Hhad}) contributes and
                  the amplitude is proportional to $a_1$; e.g.
                  $\overline B^0 \to D^+\pi^-$.
   \item[Class 2] Only the second term in (\ref{eq:Hhad}) contributes and
                  the amplitude is proportional to $a_2$; e.g.
                  $\overline B^0 \to D^0 \pi^0$. Sometimes called
                  `color-suppressed' decays.
   \item[Class 3] Both terms in (\ref{eq:Hhad}) contribute and the
                  amplitude contains both $a_1$ and $a_2$; e.g.
                  $B^- \to D^0 \pi^-$.
\end{description}
Some comments are in order. If both final-state particles are charged, then
it is Class 1, if both are neutral, then it is Class 2, if
one is neutral and the other is charged, then it depends. In $\overline B^0
\to D^+\pi^-$, the current $B\to D$ emits a $\pi$ and thus the pion decay
constant $f_\pi$ is involved. In $\overline B^0 \to D^0 \pi^0$, the current
$B\to\pi$ emits a $D$ meson and thus the $D$ meson decay constant $f_D$ is
involved. In $B^- \to D^0 \pi^-$, a class 1 amplitude and a class 2
amplitude interfere and thus both $f_\pi$ and $f_D$ are involved. Also,
note that in $\overline B^0 \to D^0 \pi^0$, the `color transparency'
argument does not apply since the color-singlet pair passing through the
cloud is now $c \overline u$ pair which are moving quite slowly, and it may
form a $D$ meson before leaving the cloud. Thus, factorization may not be a
good assumption in this case.

Heavy mesons may also decay through valence quark annihilation or
$W$-exchange processes \cite{annexch} as shown in Figure~\ref{fg:annexch}
\begin{figure}
  \vspace{3.5in}
  \caption{The annihilation and $W$-exchange processes.}
  \label{fg:annexch}
\end{figure}
which are also mediated
by interactions of types ${\bf O}_{1,2}$. Such processes have been discussed
in the context of the lifetime difference between $D^+$ and $D^0$, but
thought to be helicity-suppressed \cite{helicity-sup}, and also suppressed
by form factor effect when two-body decays are considered \cite{form-sup}.
It was suggested, however, that the helicity suppression may be lifted when
soft gluon effects are taken into account \cite{helicity-OK}. Even though
annihilation/exchange processes are usually ignored in $B$ decays, it has
not been proven that they do not significantly contribute in all types of
decays.

\subsection{{\it Experimental Test of Factorization}}

The decays $B\to PP,PV$ have definite final spin state, where $P$ is a
pseudo scalar meson and $V$ a vector meson, thus the decay rate is the
only dynamical parameter that can be tested. On the other hand, the
decays $B\to VV$ has three possible helicity amplitudes which can also be
compared against prediction of factorization.

For the test of decay rates, we take $\overline B^0 \to D^{*+} X^-$ with
$X^-$ being $\pi^-$, $\rho^-$, or $a_1^-$. As described above,
factorization allows us to estimate the decay rates of these modes from
the $q^2$-dependent form factors of the corresponding semileptonic decay
$\overline B^0\to D^{*+}l^-\nu$. In other words, there is a simple relation
between the differential decay rate of the semileptonic mode at $q^2=m_X^2$
and the corresponding non-leptonic decay rate, which can be conveniently
written as \cite{BjorkenFact}
\begin{equation}
    R\stackrel{\rm def}{\equiv}
      \frac{Br(\overline B^0 \to D^{*+} X^-)}
           { \left.\frac{dBr}{dq^2}(\overline B^0\to D^{*+}l^-\nu)
             \right|_{q^2=m_X^2} } =
      6\pi^2 f_X^2 |V_{ud}|^2  \label{eq:Rfact}
\end{equation}
where $f_X$ is the decay constant of the meson $X$. No QCD correction is
included in the expression on the right hand side \cite{NoQCD}. If QCD
correction is to be included, a reasonable choice would be to add $(C_1 +
C_2/3)^2$ to the right hand side of (\ref{eq:Rfact}).
This is because, in (\ref{eq:O12Fierz}), the contribution from the
octet current has been shown to be suppressed in the decays in question as
shown by Dugan and Grinstein \cite{Dugan-Grinstein}. However, $C_1 +
C_2/3$ is unity to the first order due to the relation $C_+^2 C_1=1$; thus,
we will proceed without QCD correction. The above formula is applicable for
$X$ being any spin-1 particle or any light spin-0 particle (assuming
factorization, of course) \cite{RosnerfX}. When the particle $X$ is spin-0,
it cannot  replace all the helicity degrees of freedom of the $D^*$
appearing in the semileptonic decay, and the formula is correct only in the
limit of $m_X\ll m_B$. The correction for pion, however, is negligible
($\sim$ 0.5\%). If $X$ is spin-1, then no such restriction applies. If
$D^*$ is replaced by $D$, then a similar helicity projection factor should
be included.

The procedure of the test is to measure the decay rate $B\to D^*X$ and
the differential semileptonic rate $d\Gamma/dq^2$ to obtain the ratio
$R$, and then compare it to the value expected from factorization:
$6\pi^2 f_X^2 |V_{ub}|^2$. The $q^2$ distribution of the semileptonic
decay $\overline B\to D^{*+} l^- \nu$ is shown in Figure~\ref{fg:q2slep},
which is a combination of ARGUS \cite{ARGUSq2} and CLEO \cite{CLEOq2} data.
\begin{figure}
  \vspace{3.5in}
  \caption{The distribution of the lepton-neutrino invariant mass ($q^2$)
for the process $\overline B\to D^{*+} l^-\nu$ as measured by CLEO and
ARGUS.}
  \label{fg:q2slep}
\end{figure}
The shape is fit to three different models
\cite{WSBsemilep,ISGWsemilep,KSsemilep} to obtain the value at given
$q^2$.

The decay constants can be obtained by the leptonic decay rate
\cite{decayconst}
\begin{equation}
     \Gamma(\pi^-\to \mu^-\overline\nu_\mu)
      = {G_F^2 f_\pi^2 \over 8\pi} m_\pi m_\mu^2
     \left(1-{m_\mu^2\over m_\pi^2}\right)^2 ,
     \label{eq:DClep}
\end{equation}
for pion which gives $f_\pi = 132$ MeV. From the tau decay rates
\begin{equation}
     \Gamma(\tau^-\to V^-\overline\nu_\tau)
       = {m_\tau^3\over 16\pi} G_F^2|V_{ud}|^2 f_V^2
     \left(1- {m_V^2\over m_\tau^2}\right)^2
     \left(1+2{m_V^2\over m_\tau^2}\right)
     \label{eq:DCtau}
\end{equation}
where $V$ is a vector or axial vector, we get $f_\rho = 197\pm3$ MeV and
$f_{a_1} = 178\pm28$ MeV. Including the effect of decay width of
meson \cite{PhamVu}, these go up to $f_\rho = 210\pm3$ MeV and $f_{a_1} =
201\pm32$ MeV. Using the isospin symmetry relation $f_{\rho^-}=f_{\rho^0}$,
the decay constant of $\rho$ can also be obtained from $\Gamma(\rho^0\to
e^+e^-)$ measured in $e^+e^-\to\rho^0$ by \begin{equation}
    \Gamma(V^0\to e^+e^-) = {4\pi\alpha^2 \over 3 m_V}c_V f_V^2
     \label{eq:DCgamee}
\end{equation}
where $c_V=1/2$ for $\rho$ \cite{cVgamee}; this gives $f_\rho = 216\pm5$
MeV which we will use.

\begin{table}[b]
\begin{center}
\begin{tabular}{|c|c c c|c c|}
\hline
\hline
    $X$   & $Br(D^{*+}X^-)$ & $dBr/dq^2$   & $R$(measured)
          & $f_X$           & $R = 6\pi^2|V_{ud}|^2f_X^2$ \\
          &   (\%)$^{(a)}$  & (\%/GeV$^2$) & (GeV$^2$)
          &   (GeV)         & (GeV$^2$)    \\
\hline
    $\pi$  & 0.265$\pm$0.036  & 0.23$\pm$0.05 & 1.15$\pm$0.30
           & 0.132$\pm$0.0005 & 0.98$\pm$0.01   \\
    $\rho$ & 0.735$\pm$0.106  & 0.25$\pm$0.04 & 2.94$\pm$0.63
           & 0.216$\pm$0.005  & 2.63$\pm$0.12    \\
    $a_1$  & 1.32 $\pm$0.30$^{(b)}$   & 0.32$\pm$0.04 & 4.13$\pm$1.07
           & 0.201$\pm$0.032   & 2.27$\pm$0.72   \\
\hline
\end{tabular}
\end{center}
 (a) The errors are statistical only.

	(b) It is assumed that $D^{*+}a_1^-$ dominates $D^{*+}\pi^+\pi^-\pi^-$
mode where the 3$\pi$ mass is between 1.0 and 1.6 GeV.

\caption{Test of factorization. Branching fraction of $B\to D^*X$ is
compared to the corresponding semileptonic decay evaluated at
$q^2=m_X^2$.} \label{tb:BDXfact} \end{table}

Table~\ref{tb:BDXfact} summarizes the result of the comparison. Note that
in taking the ratio (\ref{eq:Rfact}), uncertainty in $D^*$ detection
efficiency is canceled. This of course assumes that same $D^*$
and $D$ branching ratios are used for the measurements of $D^*l\nu$ mode and
$D^*X$ mode; a correction has been made to the values of
Figure~\ref{fg:q2slep} using the new measurements from CLEO
\cite{CLEODBrs,CLEOD*Brs}. The agreement is quite good for $\pi$ and $\rho$.
For $a_1$, the measured $R$ is about a factor of two larger than the
expected value, but statistically it is only 1.5 sigma's. This could
well be due to breakdown of factorization at $a_1$ mass of 1.26 GeV. The
branching ratio of $D^*a_1$ is determined assuming that the
$D^{*+}\pi^+\pi^-\pi^-$ final state with $1.0<M_{3\pi}<1.6$ GeV is
dominated by $a_1$. Figure~\ref{fg:D*a1} \begin{figure}
  \vspace{3.5in}
  \caption{The $3\pi$ mass distribution in the decay
    $\overline B^0 \to D^{*+}\pi^+\pi^-\pi^-$. (a) Monte Carlo
    simulation for $\overline B^0 \to D^{*+}a_1^-,a_1^-\to\rho^0\pi^-$.
    (a) Monte Carlo simulation where $\rho^0\pi^-$ is uniform in phase
    space. (c) Data with $B$-mass side bands subtracted.}
  \label{fg:D*a1}
\end{figure}
shows the $3\pi$ invariant mass distribution for the decay mode $\overline
B^0 \to D^{*+}\pi^+\pi^-\pi^-$. The $a_1$ peak is clearly seen, and amount
of non $a_1$ contribution is quite small.

As stated earlier, for $B\to VV$ decays there are helicity degrees of
freedom which cannot be uniquely determined by kinematics. The
factorization assumption leads to specific prediction for helicity
amplitudes which can then be tested experimentally. For example, once the
matrix element is factorized as
\begin{equation}
   Amp(B\to D^*\rho)\propto
     \langle \rho | (\overline u d)_{\rm had}^\mu | 0 \rangle
     \langle D^* | (\overline c b)_{{\rm had}_\mu} | \overline B \rangle,
\end{equation}
then by Lorentz invariance the rho production part $\langle \rho |
(\overline u d)_{\rm had}^\mu | 0 \rangle$ is proportional to the $\rho$
polarization vector $\epsilon^\mu$ [see (\ref{eq:dconvec})]. It then
acts the same way as the polarization vector of $W$ in semileptonic decay
resulting in the same $\rho$ polarization as that of the $W$ in
semileptonic decay at $q^2=m_\rho^2$. If factorization is not valid, this
argument cannot hold, and thus it serves as a good check of factorization.

The polarization of $\rho$ can be measured by the distribution of
$\rho^-\to\pi^0\pi^-$ polar decay angle $\theta_\rho$ in the $\rho$ rest
frame with respect to the $D^*$ direction in the same frame. Longitudinal
polarization (helicity=0) would have $\cos^2\theta_\rho$ distribution while
transverse polarization (helicity=$\pm$1) would have $\sin^2\theta_\rho$
distribution. Or equivalently, one can measure the decay angle of $D^*$
($\theta_D$) in the same way since the helicity of $D^*$ is the same as that
of $\rho$. In fact, the angular distribution can be written as
\begin{equation}
     {d\Gamma\over d\cos\theta_\rho d\cos\theta_D} \propto
      {1-a_L\over4}\sin^2\theta_\rho\sin^2\theta_D +
      a_L \cos^2\theta_\rho\cos^2\theta_D
\end{equation}
where $a_L$ is the fraction of longitudinal polarization
\begin{equation}
    a_L =           {|H_0|^2 \over |H_+|^2 + |H_0|^2 + |H_-|^2 }
\end{equation}
with $H_{+,0,-}$ being the three helicity amplitudes. Figure~\ref{fg:D*rho}
\begin{figure}
  \vspace{3.5in}
  \caption{The angular distributions for $D*$ decay angle (a) and the
       $\rho$ decay angle (b) in $\overline B^0\to D^{*+}\rho^-$.}
  \label{fg:D*rho}
\end{figure}
shows the distributions of $\theta_\rho$ and $\theta_D$ for data. A
simultaneous fit to the two angles gives
\begin{equation}
    a_L = 0.93\pm0.05\pm0.05 \qquad \hbox{(CLEO)}
\end{equation}
where the first error is statistical and the second systematic which
includes uncertainty in background subtraction and detection efficiencies.
The experimental measurement of the polarization in the semileptonic decay
is unfortunately not available at this point, and we have to compare the
above measurement to absolute theoretical prediction which requires some
assumption on form factors. One estimate using HQET \cite{RosnerfX} gives
\begin{equation}
  a_L = 0.88 \qquad \hbox{(factorization + HQET)}
\end{equation}
which is in agreement with the data.

\vskip 2.5in

This helicity=0 dominance can be intuitively understood as follows (see
the figure above). When the $\overline u d$ pair is emitted, they are
nearly collinear, and the helicities are left-handed for the $d$ quark and
right-handed for the $\overline u$ quark. Therefore the total helicity is
zero which is transferred to the final $\rho$ meson assuming that there is
no final state interaction that changes the spin state. This feature is
independent of specific choice of form factors, while it does assume
factorization.

\subsection{{\it Extraction of a$_1$ and a$_2$}}

In this section, we will take the coefficients $a_1$ and $a_2$ as free
parameters in the factorized effective Hamiltonian (\ref{eq:Hhad}), and
try to find their values by fitting to measured branching ratios. First,
we will use $B\to D\pi$ and $\psi K$ decays to demonstrate the procedure,
then a global fit to clean modes will be performed.

In order to extract $a_1$ and $a_2$, we need the form factors of $B\to D$
transition. This is quite well known; we will use the result of the fit to
the universal form factor under the framework of HQET. The relevant value
here is $F_0(q^2=m_\pi^2) = 0.58$. For the $B\to\pi$ or $K$ transition,
there is no experimental data, and a model calculation is used where the
overlap of $B$ and the light meson wave functions is obtained by
relativistic harmonic oscillator model and the $q^2$ dependence is given by
pole dominance. The coefficients of $a_{1,2}$ below are taken from
Reference~\cite{fact-rev}.

Class-I (determination of $a_1$): The decay amplitude of $\overline B^0 \to
D^+\pi^-$ (or for any two-body decay $B\to PP$) is given by
\begin{equation} \Gamma = {p\over8\pi M_B^2}|Amp|^2
\end{equation}
where $p$ is the momentum in the $B$ rest frame. Using the factorized
amplitude (\ref{eq:BDpiamp}) together with  $V_{cb}=0.045$, $V_{ud}=0.975$,
$G_F = 1.166\times10^{-5}$ (GeV$^{-2}$), $F_0(m_\pi^2) = 0.58$, and
$\tau_B=1.18$ ps, we get
\begin{equation}
Br(D^+\pi^-) = 0.264 a_1^2 \quad(\%).
\end{equation}
The measured branching ratio is $Br(D^+\pi^-)=0.29\pm0.04\%$ from CLEO,
where the error is statistical only. It then gives $a_1 = 1.1$.

Class-II (determination of $a_2$): In $\overline B^0\to D^0\pi^0$, $D$ meson
is emitted and the transition is from $B$ to $\pi$. Proceeding the same way
as before, we get
\begin{equation}
Br(D^0\pi^0) = 0.201 \left({f_D({\rm GeV})\over0.22}\right)^2 a_2^2
    \quad(\%).
\end{equation}
where the isospin factor 1/2 is included ($\pi^0$ is half $\overline u u$
and half $\overline d d$). Experimentally, only upper limit exists for this
mode: a recent number from CLEO is $Br(D^0\pi^0) < 0.035\% (90\%
C.L.)$, which corresponds to $|a_2|<0.4$.

The decay $B^-\to \psi K^-$ is also a Cabbibo-favored Class-II decay.  The
$M_B$ distribution by CLEO is shown in Figure~\ref{fg:psiK}(a) together
with other related modes that are used in the combined fit later; there are
about 60 signal events with little background. Mass peaks for $B^-\to \psi
K^-$ and $\psi K^{*0}$ by CDF are shown in Figure~\ref{fg:CDFpsiK}
\cite{CDFpsiK}.
\begin{figure}
  \vspace{3.7in}
  \caption{The beam-constrained masses for $B^-\to \psi K^-$(a), $B^0\to
   \psi K^0$(b), $B^-\to \psi K^{*-}$(c), and $B^0\to\psi K^{*0}$(d) (CLEO
    collaboration)}
  \label{fg:psiK}
\end{figure}
\begin{figure}
  \vspace{3.2in}
  \caption{Invariant mass peaks for $B^-\to \psi K^-$(a) and $B^0\to
          \psi K^{*0}$(b) by the CDF collaboration.}
  \label{fg:CDFpsiK}
\end{figure}
The transition is $B\to K$
and $\psi$ is emitted. The decay constant of $\psi$ can be obtained from
its $e^+e^-$ width: $f_\psi = 384\pm14 MeV$.
The expected branching ratio is \begin{equation}
   Br(\psi K^-) = 1.819 a_2^2 \quad(\%)
\end{equation}
where the large coefficient is primarily due to the large decay constant
of $\psi$. The measurement $Br(\psi K^-) = 0.110\pm0.015$
(CLEO) gives $|a_2| = 0.26$. One point of caution is that $a_2$ in $b\to
c\overline cs$ transition is likely to be different from $a_2$ in $b\to
c\overline u d$ transition. In fact, the values of $C_{1,2}$ themselves are
expected to be different as seen in (\ref{eq:C12bcsc}). Nonetheless, they
are often assumed to be the same and we will proceed with this assumption
for now.

Class-III (determination of $a_2/a_2$): As stated earlier, for Class-II
and Class-III decays, the factorization assumption is not well founded.
However, if we assume the factorized Hamiltonian (\ref{eq:Hhad}), we can
obtain the sign as well as the absolute value of $a_2/a_1$ through the
interference of the two types of diagrams shown in Figure~\ref{fg:BDpi}.
For example, the branching fraction of $B^-\to D^0 \pi^-$ (normalized to
$\overline  B^0\to D^+ \pi^-$) is given by \begin{equation}
   {Br(D^0 \pi^-) \over Br(D^+ \pi^-)} =
     \left[ 1 + 1.230{a_2\over a_1}
        \left({f_D(MeV)\over220}\right) \right]^2.
\end{equation}
The ratio measured by CLEO is $1.84\pm0.24\pm0.29$, and this leads to
$a_2/a_1=0.29\pm0.11$. The positive sign is a direct consequence of
$Br(D^0 \pi^-)>Br(D^+ \pi^-)$.

Tables~\ref{tb:Class-I}-\ref{tb:Class-III} summarize measurements and
expected branching ratios from the factorization model as calculated in
Reference~\cite{fact-rev}. The agreements are excellent in all cases.

\begin{table}
\begin{center}
\begin{tabular}{|c|c c|c c|}
\hline
\hline
     $\overline B^0$ & CLEO (\%)
          & ARGUS (\%)
          & Model (\%) \cite{fact-rev}
          & $a_1=1.15$
          \\
\hline
  $D^+\pi^-$      & $0.29\pm0.04\pm0.03\pm0.05^a$ &$0.48\pm0.11\pm0.11^d$
                  & $0.264a_1^2$      &  0.35    \\
  $D^+\rho^-$     & $0.81\pm0.11\pm0.12\pm0.13^a$ &$0.9\pm0.5\pm0.3^d$
                  & $0.621a_1^2$     &  0.82    \\
  $D^{*+}\pi^-$   & $0.26\pm0.03\pm0.03\pm0.01^a$ &$0.28\pm0.09\pm0.06^d$
                  & $0.254a_1^2$     &  0.34    \\
  $D^{*+}\rho^-$  & $0.74\pm0.10\pm0.13\pm0.03^a$ &$0.7\pm0.3\pm0.3^d$
                  & $0.702a_1^2$     &  0.93    \\
  $D^{*+}a_1^{-e}$   & $1.26\pm0.20\pm0.14\pm0.04^a$ &
                  & $0.97a_1^2(f_{a1}/0.22)^2$  & 1.28 \\
  $D^{**+}_{(2460)}\pi^-$& $<0.18^a$                 &
                      &   &  \\
\hline
  $D^+D_s^-$          & $1.2\pm0.7^b$    & $1.7\pm1.3\pm0.6^c$
                      & $1.213a_1^2(f_{Ds}/0.28)^2$ & 1.60    \\
  $D^+D_s^{*-}$       &                  & $2.7\pm1.7\pm0.9^c$
                      & $0.859a_1^2(f_{Ds*}/0.28)^2$ & 1.14   \\
  $D^{*+}D_s^-$       & $2.4\pm1.4^b$    & $1.4\pm1.0\pm0.3^c$
                      & $0.824a_1^2(f_{Ds}/0.28)^2$  & 1.09   \\
  $D^{*+}D_s^{*-}$    &                  & $2.6\pm1.4\pm0.6^c$
                      & $2.203a_1^2(f_{Ds*}/0.28)^2$ & 2.91   \\
\hline
\hline
  $B^-$ & & & & \\
\hline
  $D^0D_s^-$          & $2.9\pm1.3^b$    & $2.4\pm1.2\pm0.4^c$
                      & $1.215a_1^2(f_{Ds}/0.28)^2$ & 1.61    \\
  $D^0D_s^{*-}$       &                  & $1.6\pm1.2\pm0.3^c$
                      & $0.862a_1^2(f_{Ds*}/0.28)^2$ & 1.14   \\
  $D^{*0}D_s^-$       &                  & $1.3\pm0.9\pm0.2^c$
                      & $0.828a_1^2(f_{Ds}/0.28)^2$  & 1.10   \\
  $D^{*0}D_s^{*-}$    &                  & $3.1\pm1.6\pm0.5^c$
                      & $2.206a_1^2(f_{Ds*}/0.28)^2$ & 2.92   \\
\hline
\end{tabular}
\end{center}
a. Preliminary result to be submitted to Phys. Rev. D. The first error is
statistical, the second systematic, and the third error is due to
uncertainties of $D$ branching ratios.

b. Reference~\cite{CLEO_DDs}, $Br(D_S^+ \to \phi\pi^+) = 2\%$ is used.

c. Reference~\cite{ARGUS_DDs}, $Br(D_S^+ \to \phi\pi^+) = 2.7\%$ is used.

d. Reference~\cite{ARGUSmasses}.

e. All events with $3\pi$ mass between 1.0 and 1.6 GeV (after background
subtraction) are assumed to be $a_1$.
\caption{Class-I Branching ratios}
\label{tb:Class-I} \end{table}

\begin{table}
\begin{center}
\begin{tabular}{|c|c c|c c|}
\hline
\hline
     $\overline B^0$ & CLEO (\%)
          & ARGUS (\%)
          & Model (\%) \cite{fact-rev}
          & $a_1=0.26$
          \\
\hline
  $D^0\pi^0$          & $<0.035^a$   &
                      & $0.201a_2^2(f_D/0.22)^2$ & 0.014  \\
  $D^0\rho^0$         & $<0.042^a$   &
                      & $0.136a_2^2(f_D/0.22)^2$ & 0.009  \\
  $D^{*0}\pi^0$       & $<0.072^a$   &
                      & $0.213a_2^2(f_{D*}/0.22)^2$ & 0.014 \\
  $D^{*0}\rho^0$      & $<0.092^a$   &
                      & $0.223a_2^2(f_{D*}/0.22)^2$ & 0.015 \\
  $D^0\eta$           & $<0.075^a$  & & & \\
  $D^0\eta'$          & $<0.074^a$  & & & \\
  $D^0\omega$         & $<0.048^a$  & & & \\
  $D^{*0}\eta$        & $<0.086^a$  & & & \\
  $D^{*0}\eta'$       & $<0.36^a$   & & & \\
  $D^{*0}\omega$      & $<0.13^a$   & & & \\
\hline
  $\psi \overline K^0$   &$0.075\pm0.024\pm0.008^a$&$0.08\pm0.06\pm0.02^b$
                      & $1.817a_2^2$   &  0.123   \\
  $\psi \overline K^{*0}$&$0.169\pm0.031\pm0.018^a$&$0.11\pm0.05\pm0.02^b$
                      & $2.927a_2^2$   &  0.198   \\
  $\psi'\overline K^0$   &$<0.08^a$                &$<0.28^b$
                      & $1.065a_2^2$   &  0.072   \\
  $\psi'\overline K^{*0}$&$<0.19^a$                &$<0.23^b$
                      & $1.965a_2^2$   &  0.133   \\
  $\chi_{c1}\overline K^0$          & $<0.27^a$  & & & \\
  $\chi_{c1}\overline K^{*0}$       & $<0.21^a$  & & & \\
\hline
\hline
$B^-$  & & & & \\
\hline
  $\psi K^-$   &$0.110\pm0.015\pm0.009^a$&$0.07\pm0.03\pm0.01^b$
                      & $1.819a_2^2$   &  0.123   \\
  $\psi K^{*-}$&$0.178\pm0.051\pm0.023^a$&$0.16\pm0.11\pm0.03^b$
                      & $2.932a_2^2$   &  0.198   \\
  $\psi' K^-$   &$0.061\pm0.023\pm0.015^a$&$0.18\pm0.08\pm0.04^b$
                      & $1.068a_2^2$   &  0.072   \\
  $\psi' K^{*-}$ & $<0.30^a$              &$<0.49^b$
                      & $1.971a_2^2$   &  0.133   \\
  $\chi_{c1}K^-$      & $0.097\pm0.040\pm0.009^a$ & & & \\
  $\chi_{c1}K^{*-}$   & $<0.21^a$                 & & & \\
\hline
\end{tabular}
\end{center}
a. Preliminary result to be submitted to Phys. Rev. D.

b. Reference~\cite{ARGUSmasses}. Modes involving a $K_s$ are multiplied by
two to obtain the branching ratios for $\overline K^0$.
\caption{Class-II Branching ratios} \label{tb:Class-II}
\end{table}

\begin{table}
\begin{center}
\begin{tabular}{|c|c c|c c|}
\hline
\hline
     $B^-$ & CLEO (\%)
          & ARGUS (\%)
          & Model (\%) \cite{fact-rev}
          & $a_1=1.15$   \\
     & & & & $a_2=0.26$  \\
\hline
  $D^0\pi^-$    & $0.55\pm0.04$       &$0.20\pm0.08\pm0.06^b$
        & $0.265(a_1+1.230a_2(f_D/0.22))^2$          &  0.57   \\
                & $\pm0.03\pm0.02^a$  & & & \\
  $D^0\rho^-$   & $1.35\pm0.12$       &$1.3\pm0.4\pm0.4^b$
        & $0.622(a_1+0.662a_2(f_D/0.22))^2$         &  1.09   \\
                & $\pm0.12\pm0.04^a$  & & & \\
  $D^{*0}\pi^-$ & $0.49\pm0.07$       &$0.40\pm0.14\pm0.12^b$
        & $0.255(a_1+1.292a_2(f_{D*}/0.22))^2$       &  0.56   \\
                & $\pm0.06\pm0.03^a$  & & & \\
  $D^{*0}\rho^-$& $1.68\pm0.21$       &$1.0\pm0.6\pm0.4^b$
        & $0.703[a_1^2+0.635a_2^2(f_{D*}/0.22)^2$          &  1.27 \\
   & $\pm0.22\pm0.08^a$ & & $+1.487a_1a_2(f_{D*}/0.22)]$   &   \\
  $D^{*0}a_1^{-c}$ & $1.88\pm0.40$       &
        &                             &          \\
                & $\pm0.30\pm0.10^a$  & & & \\
  $D^{**0}_{(2420)}\pi^-$ & $0.11\pm0.05$    & & & \\
                       & $\pm0.04\pm0.03^a$ & & & \\
  $D^{**0}_{(2460)}\pi^-$  & $<0.15^a$         & & & \\
  $D^{**0}_{(2420)}\rho^-$ & $<0.14^a$         & & & \\
  $D^{**0}_{(2460)}\rho^-$ & $<0.5^a$          & & & \\
\hline
\end{tabular}
\end{center}
a. Preliminary result to be submitted to Phys. Rev. D. The first error is
statistical, the second systematic, and the third error is due to
uncertainties of $D$ branching ratios.

b. Reference~\cite{ARGUSmasses}.

c. All events with $3\pi$ mass between 1.0 and 1.6 GeV (after background
subtraction) are assumed to be $a_1$.
\caption{Class-III Branching ratios}
\label{tb:Class-III}
\end{table}

In order to obtain more accurate value for $a_1$ we fit four Class-I
modes, $\overline B^0\to D^+\pi^-, D^+\rho^-,D^{*+}\pi^-$, and
$D^{*+}\rho^-$. For $a_2$, we use the Class-II modes $\overline B^0\to\psi
K^0,\psi K^{*0}$ and $B^-\to\psi K^-,\psi K^{*-}$. The result is
\begin{equation}
   |a_1| = 1.15\pm0.04\pm0.04\pm0.09, \qquad
   |a_2| = 0.26\pm0.01\pm0.01\pm0.02
\label{eq:a1a2abs}
\end{equation}
where the first error is statistical, the second and the third are
systematic. The third error is due to the uncertainty in the ratio of
production and that of lifetimes of charged vs neutral $B$ mesons.
The relevant quantity is $(f_+\tau_+)/(f_0\tau_0)$ where $f_+,f_0$ are
the production fractions and $\tau_+,\tau_0$ are the lifetimes. This value
is sometimes assumed to be unity. A measurement from $Br(B^-\to D^{*0}
l\nu)/Br(\overline B^0 \to D^{*+}l\nu)$ \cite{CLEOftft} is
\begin{equation}
     {f_+\tau_+ \over f_0\tau_0} = 1.2\pm0.20\pm0.10\pm0.16.
   \quad {\rm (CLEO)}.
\end{equation}

For determination of $a_2/a_1$, we use the following four ratios of
branching fractions: $B(D^0\pi^-)/B(D^+\pi^-)$,
$B(D^0\rho^-)/B(D^+\rho^-)$, $B(D^{*0}\pi^-)/B(D^{*+}\pi^-)$, and
$B(D^{*0}\rho^-)/B(D^{*+}\rho^-)$ to obtain
\begin{equation}
   {a_2\over a_1} = 0.23\pm0.04\pm0.03\pm0.10
\end{equation}
where $(f_+\tau_+)/(f_0\tau_0) = 1.2$ is used and the last error is due
to the uncertainty in this quantity. The absolute value of $a_2/a_1$ is
consistent with the value obtained above which is $0.26/1.15 = 0.23$, and
the negative sign seems to be excluded. From (\ref{eq:a12}), we have
\begin{equation}
    {a_2\over a_1} = {C_1 + \xi C_2 \over C_2 + \xi C_1} \quad \to
     \quad \xi = {a_2/a_1 - C_2/C_1  \over1 - (C_2/C_1)(a_2/a_1) }.
\end{equation}
Using $C_1 = 1.11$, $C_2 = -0.26$, the negative value $a_2/a_1=-0.23$
corresponds to $\xi = 0.01$ and the positive value $a_2/a_1=0.23$
corresponds to $\xi = 0.44$. Thus, $\xi=0$ as suggested by an analysis
of charm decays \cite{BSW} seems to be excluded in the $B$ decays. However,
one has to keep in mind that in the analysis above, the factorization was
applied to questionable cases where emitted meson is heavy. Also, the
factorization is not expected to hold well for charm decays, so the
formulation using $a_{1,2}$ itself is in question in charm decays.

So far only Class-II modes observed are for $b\to c\overline cs$ only. As
one can see from the table, however, the present sensitivity is close to the
expected values for the $D^0\pi^0$ and related modes. It is likely that
these modes will be observed soon.

\subsection{{\it Final State Interaction}}

The factorization assumes that effect of final state interaction is
negligible. Therefore any test that is sensitive to final state
interaction is also a test of factorization.

One way is to perform an isospin analysis on a set of isospin-related
modes. For example, the relevant Hamiltonian for $B\to D\pi$ decays
has isospin structure $|I,I_z\rangle = |1,-1\rangle$
(i.e. $b\to c\overline u d$ - simply
a creation of $\overline ud$ pair as long as isospin is concerned).
Separating the Hamiltonian to an isospin violating part $S$ and an isospin
conserving part $h$, and treating $S$ as if it is a particle (spurion), we
can write, for example
\begin{eqnarray}
    S \overline B^0 &=& \sqrt{1\over3}|{3/2},-{1/2}\rangle -
                        \sqrt{2\over3}|{1/2},-{1/2}\rangle
            \nonumber \\
    D^+\pi^-        &=& \sqrt{1\over3}|{3/2},-{1/2}\rangle -
                        \sqrt{2\over3}|{1/2},-{1/2}\rangle .
\end{eqnarray}
Applying similar isospin decomposition to $S B^-$,
$\overline B^0\to D^0\pi^0$, and $B^-\to D^0\pi^-$, we get
\begin{eqnarray}
    Amp(S \overline B^0 \to D^+\pi^-) &=&
       {1\over3}\langle {3/2},-{1/2}|h|{3/2},-{1/2}\rangle
      +{2\over3}\langle {1/2},-{1/2}|h|{1/2},-{1/2}\rangle
                \nonumber \\
    Amp(S \overline B^0 \to D^0\pi^0) &=&
    {\sqrt2\over3}\langle {3/2},-{1/2}|h|{3/2},-{1/2}\rangle
   -{\sqrt2\over3}\langle {1/2},-{1/2}|h|{1/2},-{1/2}\rangle
                \nonumber \\
    Amp(S B^- \to D^0\pi^-) &=&
       \langle {3/2},-{3/2}|h|{3/2},-{3/2}\rangle
\end{eqnarray}
Because of isospin invariance of $h$, the matrix elements depend only on
the magnitude of the isospin:
$\langle {3/2},-{3/2}|h|{3/2},-{3/2}\rangle =
 \langle {3/2},-{1/2}|h|{3/2},-{1/2}\rangle$ which we define to
be $\sqrt3 A_{3\over2}$. Together with a definition $\langle
{1/2},-{1/2}|h|{1/2},-{1/2}\rangle = \sqrt{3\over2}
A_{1\over2}$, we then obtain
\begin{eqnarray}
 Amp(D^+\pi^-) &=& \sqrt{1\over3}A_{3\over2} + \sqrt{2\over3}A_{1\over2}
          \nonumber \\
 Amp(D^0\pi^0) &=& \sqrt{2\over3}A_{3\over2} - \sqrt{1\over3}A_{1\over2}\\
 Amp(D^0\pi^-) &=& \sqrt{3}A_{3\over2} \nonumber
\end{eqnarray}
where $A_{3\over2}$ and $A_{1\over2}$ are the isospin 3/2 and 1/2
amplitudes respectively. There are three unknown parameters:
$|A_{3\over2}|$, $|A_{1\over2}|$, and
$\delta=\arg(A_{3\over2}/A_{1\over2})$. Since there are three measurements
of decay rates, one can solve for the three unknowns. Then the non-zero
phase $\delta$ signifies the existence of final state interaction.
Unfortunately, the $D^0\pi^0$ mode is not observed yet at this point; we
expect, however, that it will be observed sometime soon as mentioned
earlier.

One could go further along this line if one is bold enough. One can set
$\delta=0$ and recalculate the decay rates that would have been without
the final state interaction. Then those rates may be compared with what is
expected by factorization. In fact, a phenomenologically successful
analysis of charm decay was performed in such manner \cite{BSW}. However,
there is no guarantee that all the effect of final state interaction can
be taken away by this method. There may be interactions with other final
states, for example.

Another possibility is to look at the azimuthal angular distribution in
$B\to VV$ decays. Taking $B\to D^*\rho$ as an example, the angular
distribution is given by
\begin{eqnarray}
{d\Gamma\over d c_D d c_\rho d\chi} &\propto&
    (|H_+|^2+|H_-|^2)s_D^2 s_\rho^2 + 4|H_0|^2 c_D^2 c_\rho^2 \nonumber\\
  && +2{\rm Re}(H_+^*H_-)s_D^2 s_\rho^2 \cos2\chi
    +2{\rm Im}(H_+^*H_-)s_D^2 s_\rho^2 \sin2\chi \\
  &&+4{\rm Re}(H_+^*H_0 + H_-^*H_0)s_D c_D s_\rho c_\rho \cos\chi
    +4{\rm Im}(H_+^*H_0 - H_-^*H_0)s_D c_D s_\rho c_\rho \sin\chi
     \nonumber
\end{eqnarray}
where $\theta_{D,\rho}$ are the polar decay angle of $D$ and $\rho$
decays as before, and $\chi$ is the azimuthal angle between the two decay
planes. We have used a short hand: $c_D = \cos\theta_D$, $s_D =
\sin\theta_D$ etc. If there is no final-state interaction and there is no
CP violation, then all the helicity amplitudes are relatively real. The
effect of CP violation would show up as difference of angular distribution
(as well as difference in total decay rate) between $B$ and $\overline B$
decays \cite{CPangdis}. For Cabbibo-favored modes such as $D^*\rho$,
we do not expect significant CP violation. Thus, existence of terms
proportional to $\sin\chi$ or $\sin2\chi$ signals final state interaction
\cite{KornGold}. This analysis should be able to be done with dataset
presently available, but thus far not completed.

\section{Suppressed Decays}

Now we move to rare decays which are typically Cabbibo-suppressed. We
start from charm-less two-body decays.

\subsection{{\it B Decays to Two Charmless Mesons}}

Each of the processes $B^0\to K^-\pi^+$, $\pi^+\pi^-$ could proceed through
two types of diagrams: spectator and penguin (Figure~\ref{fg:Kpi-pipi}).
When there exist more than one diagram with different weak
interaction phases and different final state interaction phases (i.e.
strong interaction), there can be CP violating decay asymmetries
\cite{CPFSI} as seen below. Suppose two diagrams contribute to a decay
$B\to f$ with amplitudes $A_1$ and $A_2{\rm e}^{i\delta}$ where $A_{1,2}$
are the weak amplitudes and $\delta$ is the FSI phase difference. Since only
relative phases matter, the weak and strong phases of the first diagram are
assumed to be zero. For the corresponding $\overline B\to \overline f$
decay, the weak phase changes its sign but the strong phase does not. This
leads to a decay asymmetry:
\begin{equation}
   Amp(B\to f) = A_1 + A_2 {\rm e}^{i\delta}, \quad
   Amp(\overline B\to \overline f) = A_1^* + A_2^* {\rm e}^{i\delta}
   \quad (A_1:{\rm real})
\end{equation}
\begin{equation}
  \vspace{2in}
\end{equation}
In our case, the weak phase of each diagram is given by that of the CKM
matrix elements which multiply the entire amplitude as coefficients. Thus
we expect that there is a weak phase difference as shown in the figure.
The strong phases, however, are difficult to estimate.

\begin{figure}
  \vspace{5in}
  \caption{Diagrams that can contribute to $B\to K\pi,\pi\pi$.}
  \label{fg:Kpi-pipi}
\end{figure}

If we assume the flavor $SU(3)$ symmetry, then the ratio of amplitudes are
\begin{equation}
   \left.{K\pi \over \pi\pi}\right|_{\rm spectator} \sim \lambda \qquad
   \left.{K\pi \over \pi\pi}\right|_{\rm penguin}  \sim {1\over\lambda}
\end{equation}
where $\lambda$ is the Cabbibo suppression factor ($\sim 0.2$). It is
expected that the spectator diagram will dominate in $B^0\to\pi^+\pi^-$.
Then if there is no penguin contribution, the $K^-\pi^+$ branching ratio
should be $\lambda^2\sim0.04$ times smaller than that of $\pi^+\pi^-$.
Thus, if the rate of $K^-\pi^+$ is comparable or greater than
$\pi^+\pi^-$, then it is likely that the $K^-\pi^+$ rate is dominated by
the penguin diagram. When there is a large disparity in magnitudes of the
two diagrams, the expected CP violation will be small independent of the
phases.

One should note, however, that there is a possibility that $B\to K\pi$
can occur through final state re-scatterings. This could occur through
intermediate states involving two charmed mesons as
\begin{equation}
    B^0\to D^-D_S^+ \to K^+\pi^- , \qquad B^0\to D^-D^+ \to \pi^-\pi^+
\end{equation}
which corresponds to replacing the top quark loop in the penguin diagrams by
a charm quark which will be on-shell as shown below and can be considered
to be a dispersive version of penguin diagram.
\begin{equation}
  \vspace{1.5in}
\end{equation}
Such process will result in a large
FSI phase, and can interfere with the top quark penguin diagram to
generate a CP violation as originally postulated by Bander, Silverman and
Soni \cite{CPpenguin}.

Approximate rate of $\pi^+\pi^-$ can be estimated from the measured $B^0\to
D^-\pi^+$ rate quite reliably:
\begin{equation}
   Br(\pi^+\pi^-) \sim \left|{V_{ub}\over V_{cb}} \right|^2
   Br(D^+\pi^-) \sim
1\times10^{-5}
\end{equation}
where the effect of form factor will reduce it somewhat and that of phase
space will increase it somewhat. The estimation of the $K\pi$ rate
requires the coefficient of the penguin operator, and the uncertainty is
greater; the theoretical estimates are in the same range as the $\pi\pi$
mode \cite{Kpi-penguin}.

Experimentally, the signature on $\Upsilon(4S)$ is a rather
spectacular high-momentum back-to-back tracks of $p\sim 2.6$ GeV. This is
the maximum momentum a $B$-decay can emit and the background is
dominated by continuum events; thus, cuts are made on event shapes to
reject 2-jet like events and the fast back-to-back tracks are required not
to be aligned with the jet axis of event. For a $B\overline B$ pair event,
the event shape is spherical and there is little correlation between the
event axis and the direction of the back-to-back tracks. Then, as before,
the energy difference $\Delta E$ and the beam-constrained mass $M_B$ is used
to select the candidates [see (\ref{eq:MBdE})].

When masses are correctly assigned to the tracks, the $\Delta E$ resolution
is 25 MeV. The $dE/dx$ information in the drift chamber is
used to separate kaon and pion. The $dE/dx$ resolution is 6.5\%\ and
provides 1.8$\sigma$ $K-\pi$ separation per track. Each candidate is
assigned the most likely masses ($\pi\pi$, $K\pi$, or $KK$), then $\Delta
E$ is calculated. The beam-constrained mass, on the other hand, does not
depend on the mass assignments and the resolution is 2.5 MeV.
Figure~\ref{fg:KpiMBdE}(a) shows the $M_B$ distribution for $K\pi$ and
$\pi\pi$ candidates after 2-$\sigma$ cut on $\Delta E$ around zero. The
shaded events are the $\pi\pi$ candidates. One can see an enhancement at
the nominal $B$ mass of 5.280 GeV. The $\Delta E$ distribution after the
2-$\sigma$ cut on $M_B$ is shown in Figure~\ref{fg:KpiMBdE}(b). Again,
there is a peak around the nominal region near $\Delta E=0$.
\begin{figure}
  \vspace{3.5in}
  \caption{Sum of $K\pi$ sample and $\pi\pi$ sample.
           (a) The beam-constrained mass $M_B$ after the 2-$\sigma$ cut on
           $\Delta E$. (b) $\Delta E$ distribution
           after the 2-$\sigma$ cut on $M_B$. The shaded events are
           the events assigned to be $\pi\pi$.}
  \label{fg:KpiMBdE}
\end{figure}
For the final extraction of numbers, an un-binned maximum likelihood fit is
performed with $\Delta E$, $M_B$, $dE/dx$, and an event shape variable as
parameters. Here $\Delta E$ is calculated assuming $\pi\pi$. The result is
shown in Table~\ref{tb:Kpi-result} \cite{CLEOKpi}.
\begin{table}
\begin{center}
\begin{tabular}{|c|c c|}
\hline
\hline
    Mode & Br($10^{-5}$) & Upper Limit ($10^{-5}$)\\
\hline
    $\pi^+\pi^-$ &  $1.3^{+0.8}_{-0.6}\pm0.2$ & 2.9   \\
    $K^+\pi^-$   &  $1.1^{+0.7}_{-0.6}\pm0.2$ & 2.6   \\
    $K^+K^-$     &                            & 0.7   \\
\hline
    $K^+\pi^- + \pi^+\pi^-$ & $2.4^{+0.8}_{-0.7}\pm0.2$ &  \\
\hline
\end{tabular}
\end{center}
\caption{Measured branching fractions and 90\%\ confidence level upper
       limits.} \label{tb:Kpi-result}
\end{table}

When $\Delta E$ is calculated assuming $\pi\pi$, the value shifts down by
42 MeV if the actual tracks are $K\pi$. Since the $\Delta E$ resolution is
25 MeV, this by itself can provide 1.7$\sigma$ separation between $K\pi$
and $\pi\pi$. The available particle identifications are not good
enough to cleanly separate the two. When $K\pi$ and $\pi\pi$ are combined
there is a substantial signal of about 3.5$\sigma$. The central value of
$\pi^+\pi^-$ mode is consistent with the expected value of
$1\times10^{_5}$. If we take the central value of the $K^+\pi^-$ mode at
its face value, then penguin diagrams (t-loop or the re-scattering c-loop)
are likely to be dominating the $K^+\pi^-$ mode.

\subsection{{\it b to s Radiative Decays}}

Another rare process a penguin diagram is expected to contribute is the
radiative $b\to s$ transition through emission and re-absorption of $W$.
\begin{equation}
  \vspace{1.5in}
\end{equation}
At the lowest order, the GIM suppression is operative and it depends on the
top mass ($m_t$) strongly, and $Br(B\to X_s \gamma)$ changes from
$0.5\times10^{-4}$ at $m_t=100$ GeV to $1.4\times10^{-4}$ at $m_t = 200$
GeV. With QCD correction \cite{bsgamQCD}, the GIM suppression is loosened
(`soft' GIM suppression) and as a result the rate is substantially enhanced
and becomes a slow function of $m_t$. The enhancement factor is $\sim$5 at
$m_t = 120$ GeV to give $Br(B\to X_s \gamma)\sim3.5\times10^{-4}$.
Theoretical estimate for the exclusive mode $B\to K^*\gamma$ is more
uncertain due to the unknown transition matrix element $B\to K^*$
\cite{ThBK*gam}. One estimate based on HQET gives $Br(B\to
K^*\gamma)=(1.4-4.9)\times10^{-5}$ \cite{HQBK*gam}.

The experimental signature \cite{CLEOK*gam} is a monochromatic hard photon
(2.6 GeV) recoiling against $K^*\to K\pi$ decay. We look for both $B^0\to
K^{*0}\gamma$ and $B^-\to K^{*-}\gamma$. The $K^*$'s are searched for in
the modes $K^{*0}\to K^+\pi^-$ and $K^{*-}\to K^-\pi^0,K_s\pi^-$. Again the
background is dominated by continuum events since such high-energy photon
is at the kinematic limit of $B$ decay. The continuum backgrounds are
reduced by requiring that the events be not 2-jet like and that the hard
photon be not aligned to the event axis. If the photon forms a $\pi^0$ or
$\eta$ with another photon then it is rejected. Figure~\ref{fg:BK*gam}
shows the $M_B$ distribution after the cut $|\Delta E|<90$ MeV
(2.2$\sigma$).
\begin{figure}
  \vspace{3.5in}
  \caption{The $M_B$ distribution for $B\to K^*\gamma$ after $\Delta E$
    cut.}
  \label{fg:BK*gam}
\end{figure}
\begin{figure}
  \vspace{3.5in}
  \caption{Single photon spectrum after continuum subtraction. The signal
         $b\to s\gamma$ would show up
         in the region 2.2 to 2.7 GeV.}
  \label{fg:bsgaminc}
\end{figure}
There is a clear signal observed with $6.6\pm2.8$ events in $B^0$ mode and
$4.1\pm2.3$ events in $B^-$ modes. The branching fractions are
\begin{eqnarray}
    Br(B^0\to K^{*0}\gamma) &=& (4.0\pm1.7\pm0.8)\times10^{-5}, \nonumber\\
    Br(B^-\to K^{*-}\gamma) &=& (5.7\pm3.1\pm1.1)\times10^{-5}
   \qquad ({\rm CLEO})
\end{eqnarray}
If we assume isospin symmetry, then
\begin{equation}
    Br(B^0\to K^{*0}\gamma) =  Br(B^-\to K^{*-}\gamma)
     = (4.5\pm1.5\pm0.9)\times10^{-5}
\end{equation}
which is consistent with theoretical estimates based on the standard model
where the penguin contribution dominates. Another possibility is that the
$B\to K^*\gamma$ transition may occur through $\psi K*$ by vector dominance
\cite{K*gamVD}
\begin{equation}
    B\to \psi K^* \to \gamma K^*  \qquad \hbox{(vector dominance)}.
\end{equation}
or other long distance effects \cite{K*gamLD}. Such processes have been
estimated and found to be at least an order of magnitude smaller than the
observed rate.

The inclusive transition $B\to X_s\gamma$ can be searched by looking for the
hard photon without reconstructing $X_s$ where the mass of $X_s$ lies in
the typical strange meson region (0.5 to 2 GeV). Similar cuts as before to
reduce continuum backgrounds are applied. Figure~\ref{fg:bsgaminc} shows
the continuum-subtracted (see Section 1) photon spectrum. The signal
region is around 2.2 to 2.7 GeV. There seems to be some enhancement, but
it is not statistically significant; thus, we set an upper limit
\begin{equation}
   Br(b\to s \gamma) < 5.4\times10^{-4} \qquad({\rm CLEO} \cite{bsgIncl}).
\end{equation}
Such measurement places stringent constraints on non-standard physics, in
particular two-Higgs-doublet models \cite{ChHiggs}. The W-top loop can
be replaced by loops involving charged Higgs, neutralinos, gluinos, and
squarks etc \cite{BYSMBertolini}. For example, in the minimal
supersymmetric model with two Higgs doublets, the mass of the CP-odd
neutral Higgs $A^0$ (which is related to the mass of the charged Higgs)
is ruled out for $m_{A^0}<250$ GeV.
\cite{HiggJH}.

\begin{center}
{\bf Acknowledgement}
\end{center}

The author would like to thank Mike Dugan, Giulia Ricciardi, Thorsten
Ohl, and Mitch Golden for fruitful discussions. A major credit should go to
the editing staff whose extraordinary patience made it possible that this
article be completed. This research was supported primarily by the DOE via
grant DEFG-029-1ER-40-654 and in part by NSF grant PHY-92-18167.

\pagebreak


\begin{thebibliography}{99}
\bibitem{B*corr}
   The references quote the photon energy in the $B^*$ rest frame. A
   correction of $\Delta M^2/2M_{B^*}$ is applied to convert it to the
   mass difference.
\bibitem{fact_oper}
   It is important to note that the current operator
   $(\overline q q')$ and the `factorized hadron operator'
   $(\overline q q')_{\rm had}$ are two different objects. A same current
   operator can contribute to more than one factorized hadron operators.
\bibitem{NoQCD}
   One may add $a_1^2$ to the expression, but what to take for $\xi$ is
not clear and the uncertainty is of order 10\%. One has to be careful
not to use the value of $a_1$ which is obtained by phenomenological
fit to hadronic rates together with semileptonic form factors. That
would be a biased test of factorization.
\bibitem{decayconst}
   In principle, one can make a QED radiative correction; the result,
   however, does not vary significantly for the purpose of this analysis.
   See Ref \cite{PDG}, p VII.1.
\bibitem{cVgamee}
   The parameter $c_V$ is the charge squared of the valence
   quark properly averaged over the quarks. For example, $c_V=1/9$ and
   $4/9$ for $\phi$ and $\psi$, respectively. For $\rho^0$,
   $c_V=|Q_u\langle u\overline u |\rho^0\rangle
     +Q_d\langle d\overline d |\rho^0\rangle |^2 = 1/2$, where $Q_i$ is
   the charge of quark type i.
\end{thebibliography}
\end{document}